\documentclass{ws-p10x7}

\begin{document}

\title{Bilinear R-parity violating SUSY: Solving the 
solar and atmospheric neutrino problems}

\author{\underline{M. Hirsch}, W. Porod and J. W. F. Valle}

\address{Instituto de F\'{\i}sica Corpuscular 
-- C.S.I.C., Departamento de F\'{\i}sica Te\`orica, 
Universitat de Val\`encia, Edificio Institutos de Paterna,
Apartado de Correos  22085, 46071 Val\`encia} 

\author{M. A. D\'{\i}az}

\address{Fac. de F\'\i sica, Universidad Cat\'olica de Chile, 
          Av. Vicu\~na Mackenna 4860, Santiago, Chile}

\author{J. C. Rom\~ao}

\address{Depto. de F\'\i sica, Inst. Superior T\'ecnico,
          Av. Rovisco Pais 1, $\:\:$ 1049-001 Lisboa, Portugal}

\twocolumn[\maketitle\abstract{
Bilinear R-parity violation is a simple extension of the MSSM allowing 
for Majorana neutrino masses. One of the three neutrinos picks up mass 
by mixing with the neutralinos of the MSSM, while the other two neutrinos 
gain mass from 1-loop corrections. Once 1-loop corrections are 
carefully taken into account the model is able to explain solar and 
atmospheric neutrino data for specific though simple choices of the 
R-parity violating parameters.
}]

\section{Introduction}

Atmospheric neutrino data so far provides the most compelling evidence
for non-zero neutrino masses \cite{Fukuda:1998mi}, indicating large
mixing among neutrino flavours and a typical $\Delta m^2_{ij}$ of the
order of $\Delta m^2_{ij} \sim$ (few) $10^{-3}$ eV$^2$. On the other
hand, despite new high quality data, the long standing solar neutrino
problem still allows both large and small mixing angle values for 
neutrinos \cite{update00} and $\Delta m^2 \sim {\cal O}(10^{-5})$
$eV^2$ or $\Delta m^2 \sim {\cal O}(10^{-7})$ $eV^2$ \cite{update00}
(SMA and LMA or LOW solutions of the solar neutrino problem,
respectively).

Many attempts to explain neutrino masses can be found in the
literature \cite{hepph}. Here we summarize the work of
\cite{Hirsch:2000ef}. It is based on a simple bilinear R-parity
violating (BRPV) extension of the MSSM. This model, despite being
minimalistic can explain atmospheric and solar neutrino data for
specific ranges of model parameters. Its attractiveness lies in the
fact that these parameter choices necessary to solve the neutrino
problems give at the same time definite predictions for accelerator
physics. \cite{valichep}

\section{Bilinear R-parity violating SUSY}

In the simplest extension of the MSSM including R-parity violation 
the superpotential contains just 3 additional bilinear terms \cite{drv98}
\begin{equation}
 W = W_{MSSM} +  \epsilon_i {\hat L_i} {\hat H_u}
\label{eq:pot}
\end{equation}
They violate lepton number by one unit and therefore necessarily
generate Majorana neutrino masses. Corresponding bilinear R-parity
violating terms appear in the soft SUSY breaking terms, but strictly
speaking these are not independent parameters because of the tadpole
conditions \cite{drv98}.

In this model at tree-level only one neutrino picks up a mass via 
mixing with the neutralinos. This tree-level mass can be estimated 
by \cite{Hirsch:1999kc}

\begin{equation}
m_{\nu} = 
\frac{M_1 g^2 + M_2 {g'}^2}{4 det({\cal M}_{\chi^0})} 
|{\vec \Lambda}|^2
\label{eq:treemass}
\end{equation}
Here, $M_1$ and $M_2$ are the MSSM gaugino masses, ${\cal M}_{\chi^0}$ 
is the MSSM neutralino mass matrix and ${\vec \Lambda}$ is defined by 
$\Lambda_i = \epsilon_i v_d + \mu \langle {\tilde \nu_i} \rangle$, with 
$\langle {\tilde \nu_i} \rangle$ being scalar neutrino vevs. 

Since only one neutrino gains mass at tree-level in BRPV, to study 
solar and atmospheric neutrino problems at the same time, it is 
necessary to include 1-loop corrections. Details are given in 
\cite{Hirsch:2000ef}.

\section{Numerical results}
\begin{figure}
\setlength{\unitlength}{1mm}
\begin{picture}(40,40)
\put(-15,-80)
{\mbox{\epsfig{figure=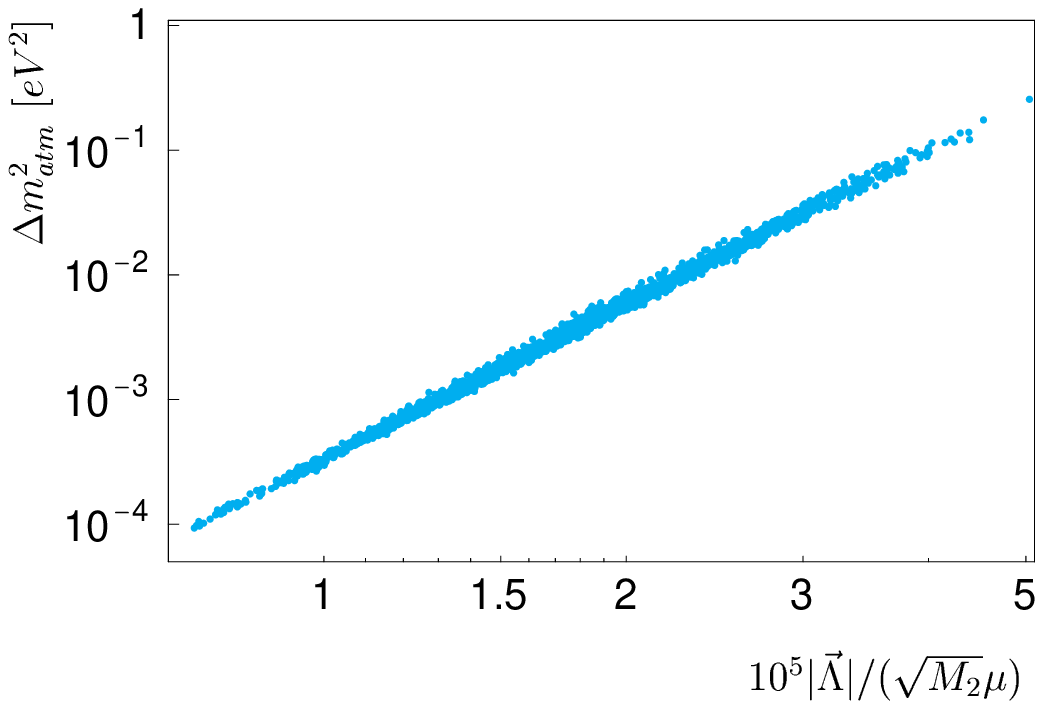,height=18.0cm,width=12.0cm}}}
\end{picture}
\caption{The atmospheric $\Delta m^2$ as a function of 
$|{\vec \Lambda}|/(\sqrt{M_2} \mu)$. The figure shows how the tree-level 
approximation can be used to fix the largest mass scale in the bilinear 
model.}
\label{fig:atmmas}
\end{figure}

After including 1-loop corrections the BRPV model produces for nearly 
all choices of parameters a hierarchical mass spectrum. The largest 
neutrino mass can then usually be estimated by the tree-level value. 
This is demonstrated in Fig. \ref{fig:atmmas}, where we show 
$\Delta m^2_{atm}$ 
as a function of $|{\vec \Lambda}|/(\sqrt{M_2} \mu)$. As the figure 
shows, correct $\Delta m^2_{atm}$ can be easily obtained by an appropriate 
choice of $|{\vec \Lambda}|$. 

\begin{figure}
\setlength{\unitlength}{1mm}
\begin{picture}(40,40)
\put(-15,-80)
{\mbox{\epsfig{figure=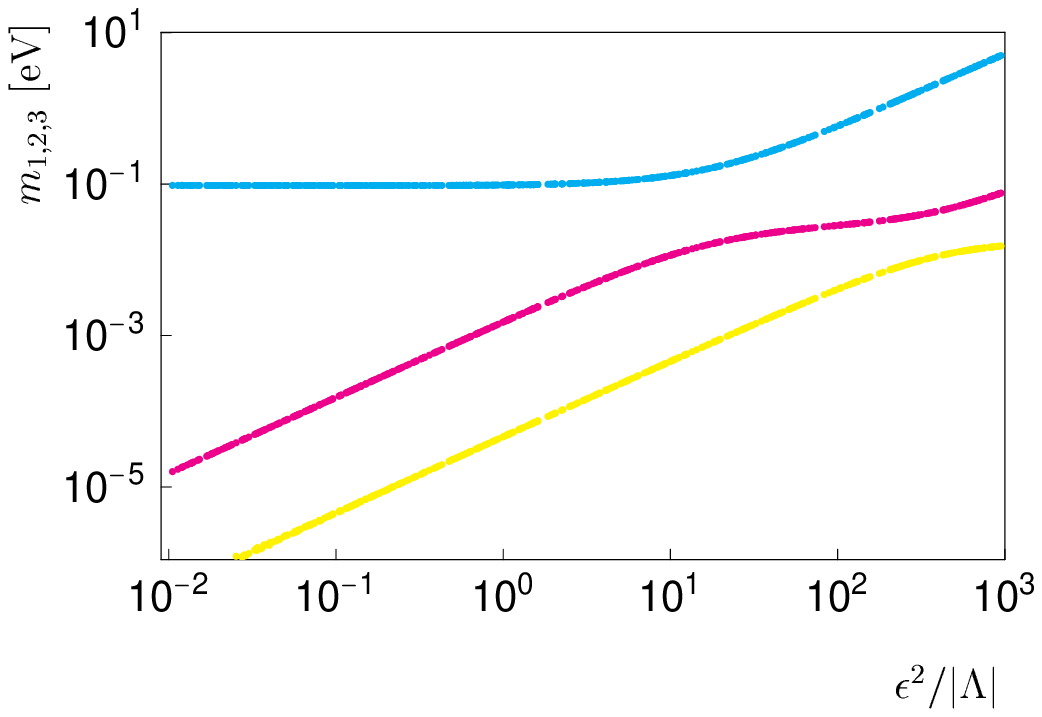,height=18.0cm,width=12.0cm}}}
\end{picture}
\caption{The solar $\Delta m^2$ as a function of 
$\epsilon^2|/{\vec \Lambda}|$ for otherwise fixed parameters of the 
model. The figure shows how the importance of loop corrections 
increases with increasing $\epsilon^2|/{\vec \Lambda}|$. }
\label{fig:solmas}
\end{figure}

The solar mass scale, on the other hand, is entirely generated at 
1-loop order and therefore depends on the model parameters in a 
complicated way. Fig. \ref{fig:solmas} shows one example. The parameter 
$\epsilon^2|/{\vec \Lambda}|$ is most important for determining the 
size of the loop corrections, but loops also show a strong dependence 
on $\tan\beta$. 

Turning to the discussion on neutrino angles, we note that as long 
as the 1-loop corrections are not larger than the tree-level contribution, 
the flavour composition of the 3rd mass eigenstate is approximately 
given as

\begin{equation}
U_{\alpha 3} \approx 
 \Lambda_\alpha / | \Lambda | .
\end{equation}

\begin{figure}
\setlength{\unitlength}{1mm}
\begin{picture}(40,40)
\put(-15,-80)
{\mbox{\epsfig{figure=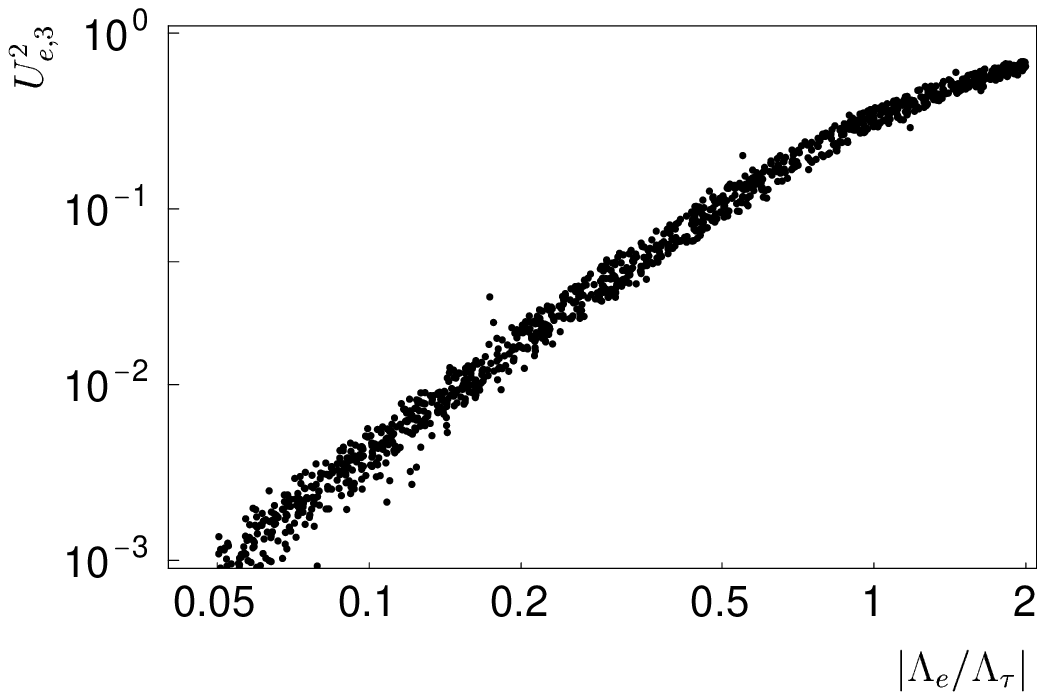,height=18.0cm,width=12.0cm}}}
\end{picture}
\caption{$U_{e3}^2$ as a function of $\Lambda_{e}/\Lambda_{\tau}$. }
\label{fig:chooz}
\end{figure}

\begin{figure}
\setlength{\unitlength}{1mm}
\begin{picture}(40,40)
\put(-15,-80)
{\mbox{\epsfig{figure=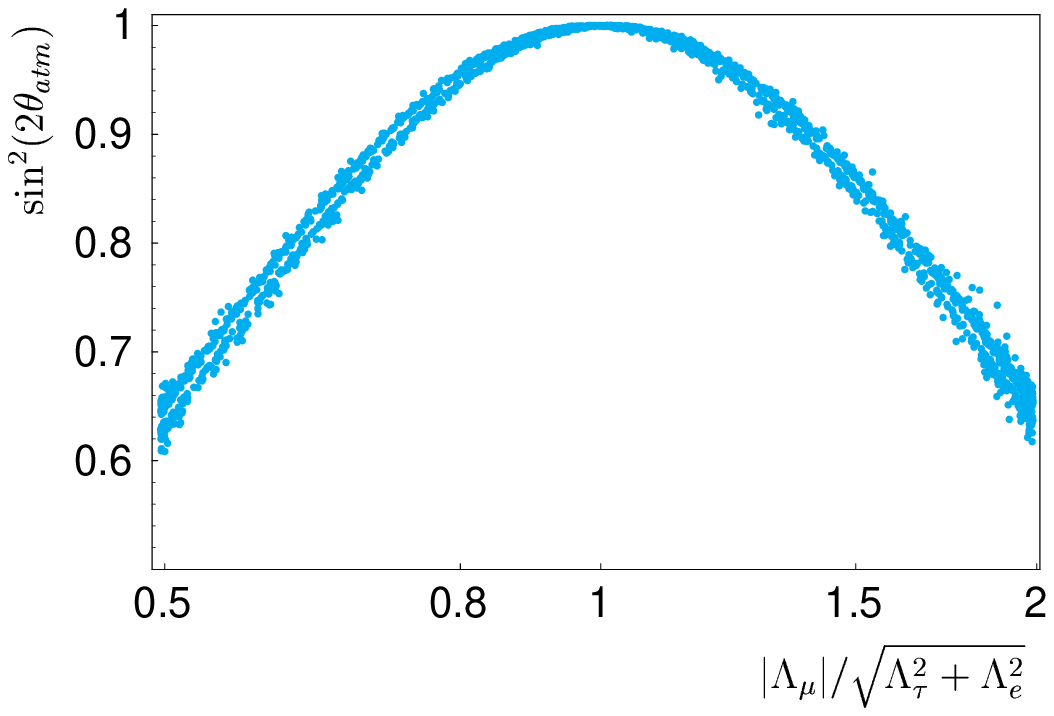,height=18.0cm,width=12.0cm}}}
\end{picture}
\caption{Atmospheric neutrino mixing angle as a function of 
$\Lambda_{\mu}/\sqrt{\Lambda_{\tau}^2+\Lambda_{e}^2}$. Since 
$\Lambda_{e} \ll \Lambda_{\mu}, \Lambda_{\tau}$ is required by 
the reactor neutrino data, $\Lambda_{\mu} \simeq \Lambda_{\tau}$ is 
needed to obtain large atmospheric neutrino mixing.}
\label{fig:atm}
\end{figure}

Since atmospheric and reactor \cite{chooz} neutrino data tell us 
that $\nu_{\mu} \rightarrow 
\nu_{\tau}$ oscillations are preferred over $\nu_{\mu} \rightarrow 
\nu_{e}$ oscillations, we conclude that $\Lambda_e \ll \Lambda_{\mu} 
\simeq \Lambda_{\tau}$ is required for BRPV to fit the data. This is 
shown in figs. \ref{fig:chooz} and \ref{fig:atm}.

\begin{figure}
\setlength{\unitlength}{1mm}
\begin{picture}(40,40)
\put(-15,-80)
{\mbox{\epsfig{figure=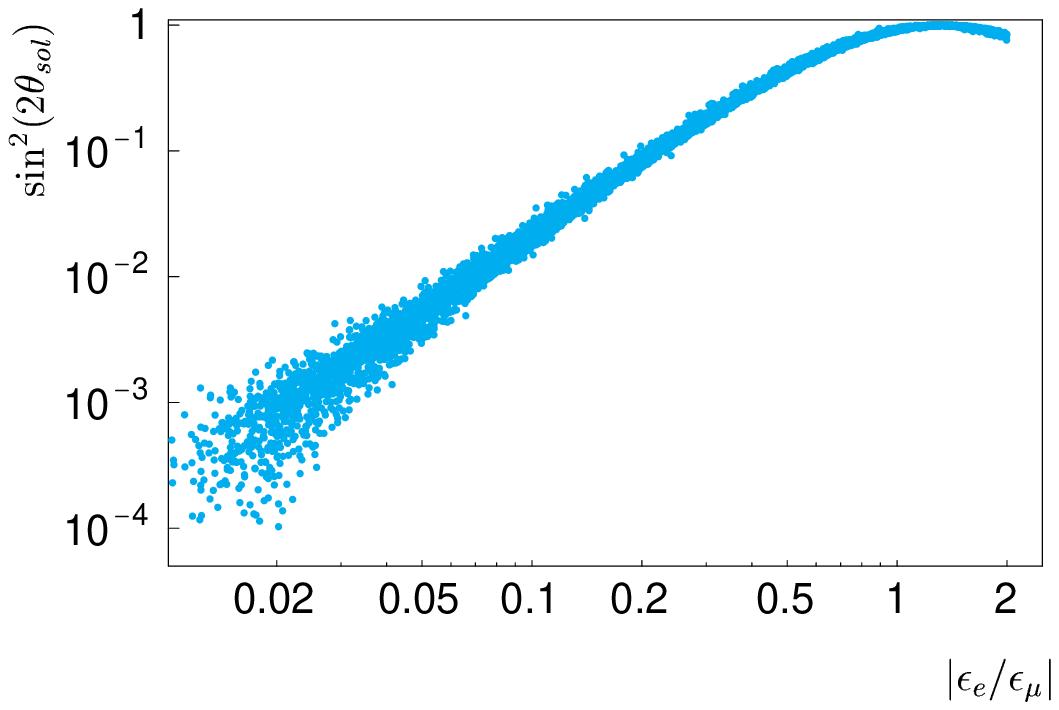,height=18.0cm,width=12.0cm}}}
\end{picture}
\caption{Solar neutrino mixing angle as a function of 
$\epsilon_{e}/\epsilon_{\mu}$.} 
\label{fig:sol}
\end{figure}

For the solar angle the situation is more complex. As explained in
\cite{Hirsch:2000ef} there are two cases to distinguish. With the
usual minimal supergravity unification assumptions, ratios of
$\epsilon_i/\epsilon_j$ fix the ratios of $\Lambda_i/\Lambda_j$. Since
atmospheric (and reactor) neutrino data imply that $\Lambda_e \ll
\Lambda_{\mu},\Lambda_{\tau}$ only the small angle solution to
the solar neutrino problem can be obtained in this case.  However,
even a very tiny deviation from universality of soft parameters at the
unification scale relaxes this constraint sufficiently, such that also
large angle solar solutions can be obtained in our model, see Fig.
\ref{fig:sol}.

\section{Conclusions}

Bilinear R-parity violating SUSY, despite being a very simple
extension of the MSSM can explain atmospheric and solar neutrino data
\cite{Hirsch:2000ef}, once 1-loop corrections are taken carefully into
account. The main attractiveness of the model, however, lies in the
fact that it can be tested at future accelerators. In \cite{valichep} we 
discuss the definite predictions made for neutralino decays.

\section*{Acknowledgments}
This work was supported by DGICYT grants PB98-0693 and SB97-BU0475382
(W.~P.), by the TMR contracts ERBFMRX-CT96-0090 and ERBFMBICT983000
(M.~H.).  

\begin{thebibliography}{99}
  
\bibitem{Fukuda:1998mi} Y.~Fukuda {\it et al.}  Phys. Rev. Lett. {\bf
    81}, 1562 (1998); and hep-ex/9805006
  
\bibitem{update00} N. Fornengo, M. Gonzalez-Garcia \&
  J. Valle, Nucl. Phys.~{\bf B580}~(2000)~58; M. Gonzalez-Garcia
  et al, Nucl. Phys. {\bf B573}, 3 (2000); an updated discussion of
  neutrino data was presented by M.~C.~Gonzalez-Garcia at this
  conference, and is avalaible from \texttt{http://neutrinos.uv.es//}.
  
\bibitem{hepph} A simple search at 
\texttt{http://xxx.lanl.gov/find/hep-ph} reveals
  hundreds of attempts to explain neutrino masses in the light of
  atmospheric and solar data

\bibitem{Hirsch:2000ef} M. Hirsch, M.A. Diaz, W. Porod, J.C. Romao
and J.W.F. Valle, hep-ph/0004115,  to appear in Phys. Rev.  {\bf D}; 
J.C. Rom\~ao, M.A. Diaz, M. Hirsch, W. Porod and J.W.F. Valle,
Phys. Rev. {\bf D61}, 071703 (2000) 

\bibitem{valichep}M. Hirsch, W. Porod, J.C. Romao and J.W.F. Valle, 
talk at this conference

\bibitem{drv98}M.A. Diaz, J.C. Romao and J.W.F. Valle, Nucl. Phys. 
{\bf B524} (1998) 60  

\bibitem{Hirsch:1999kc} M. Hirsch and J. W. F. Valle,
Nucl. Phys. {\bf B557}, 60 (1999); M.~Hirsch,
J.~C.~Romao and J.~W.~F.~Valle, Phys.\ Lett.\ {\bf B486} (2000) 255

\bibitem{chooz} M. Apollonio et al, Phys.Lett. {\bf B466} (1999) 415;
  F. Boehm et al, hep-ex/9912050

\end{thebibliography}
\end{document}